\documentclass[a4paper,fleqn,usenatbib]{mnras}
\usepackage[T1]{fontenc}
\usepackage{ae,aecompl}
\usepackage{graphicx}	
\usepackage{amsmath}	
\usepackage{amssymb}	
\newcommand{\Lya}{Ly$\alpha$}
\def\Lyb{Ly$\beta$}
\def\Lyg{Ly$\gamma$}

\def\Lyn{Ly$n$}
\def\HI{\hbox{H~$\scriptstyle\rm I$}}
\def\HII{\hbox{H~$\scriptstyle\rm II$}}

\def\nHI{{\rm HI}}
\def\nH{{\rm H}}

\def\xiunits{\,{\rm erg^{-1}\,Hz}}
\def\nunits{\,{\rm cm^{-3}\,Hz^{-1}}}

\def\emunits{\,{\rm erg\,s^{-1}\,Mpc^{-3}\,Hz^{-1}}}

\def\spose#1{\hbox to 0pt{#1\hss}}
\def\lta{\mathrel{\spose{\lower 3pt\hbox{$\mathchar"218$}}
     \raise 2.0pt\hbox{$\mathchar"13C$}}}
\def\gta{\mathrel{\spose{\lower 3pt\hbox{$\mathchar"218$}}
     \raise 2.0pt\hbox{$\mathchar"13E$}}}
\def\ni{\noindent}

\title[Global 21-cm Signal]{Constraints on Early Star Formation from the 21-cm Global Signal}   

\author[P. Madau]{Piero Madau\\
Department of Astronomy \& Astrophysics, University of California, 1156 High Street, Santa Cruz, CA 95064, USA
}

\begin{document}
\label{firstpage}
\pagerange{\pageref{firstpage}--\pageref{lastpage}}
\maketitle

\begin{abstract}
The tentative detection  by the EDGES experiment of a global 21-cm absorption trough centered at redshift 17 opens up the opportunity to study the birth
of the first luminous sources, the intensity of radiation backgrounds at cosmic dawn, the thermal and ionization history of the young intergalactic 
medium. Here, we focus on the astrophysical implications of the \Lya\ photon field needed to couple the spin temperature to the kinetic temperature of 
the gas at these early epochs. Under the basic assumption that the 21-cm signal is activated by extremely metal-poor stellar systems, we show that the 
EDGES results are consistent with an extrapolation of the declining galaxy UV luminosity density measured at $4\lta z\lta 9$ by deep {\it HST} observations. 
A substantially enhanced star formation rate density or new exotic sources of UV photons are not required at the redshifts of the EDGES signal. The 
amount of ionizing radiation produced by the same stellar systems that induce \Lya\ coupling is significant, of order 0.5 LyC photons per H-atom per 100 
Myr. To keep hydrogen largely neutral and delay the reionization process consistently with recent {\it Planck} CMB results, mean escape fractions of 
$f_{\rm esc}\lta 20$\% are required at $z>15$.
\end{abstract}


\section{Introduction}

The cosmic microwave background (CMB) spectrum is expected to show an absorption feature at frequencies below 130 MHz imprinted at cosmic dawn, 
when the universe was flooded with \Lya\ photons emitted from some of the very first stars and before it was reheated and reionized by 
Lyman-continuum (LyC) and X-ray radiation \citep{madau97,tozzi00}. The absorption signal corresponds to the redshifted 21-cm hyperfine transition 
of the ground state of neutral hydrogen and arises from the indirect coupling of the spin temperature $T_S$ to the kinetic temperature $T_K<T_{\rm CMB}$ 
of the intergalactic medium (IGM) via the Wouthuysen-Field effect. The EDGES collaboration has recently reported the detection of a flattened absorption trough 
-- centered at 78 MHz and with an amplitude of 0.5 K --  in the sky-averaged radio spectrum that places the onset of this ``\Lya-coupling era" at 
redshift $z\lta 20$, 180 Myr after the Big Bang \citep{bowman18}. The high-frequency cut-off of the EDGES absorption profile suggests that hydrogen was heated 
to above the CMB temperature less than 100 million years later, at $z\simeq 15$.
While its anomalous amplitude may indicate the need for new physics or exotic astrophysics \citep[e.g.,][]{barkana18,ewall-wice18,feng18,slatyer18}, 
we do not attempt here to explain this signal altogether. Rather, we focus on the constraints imposed by the required Wouthuysen-Field coupling strength 
on UV radiation backgrounds and the galaxy emissivity at first light. Such constraints hold under the assumption that the 21-cm signal is activated by young, metal-poor stellar 
systems regardless of the mechanisms that produce the stronger-than-expected EDGES measurement.

\section{21-cm Radiation from Cosmic Dawn}

We start by briefly reviewing the basic theory of the sky-averaged 21-cm signal from the earliest stages of the galaxy formation process 
\citep[e.g.,][]{madau97,shaver99,ciardi03,chen04,barkana05,furlanetto06,pritchard08,fialkov14,cohen17}). 
The radiative transfer equation in the Rayleigh-Jeans limit yields the observed brightness temperature relative to the CMB 
$T_{\rm 21}\equiv (T_S-T_{\rm CMB})(1-e^{-\tau})/(1+z)$, 

\begin{equation}
T_{\rm 21}(\nu_0) \simeq {3c^3 \hbar A_{10}n_{\rm HI}\over 16 k_B \nu_{10}^2H(z)(1+z)}\left(1-{T_{\rm CMB}\over T_S}\right),
\end{equation}
at the frequency $\nu_0=\nu_{10}/(1+z)$. Here, $\nu_{10}=1420.4$ MHz is the hyperfine transition frequency of atomic hydrogen, $T_S$ is the spin 
or excitation temperature, $T_{\rm CMB}=2.725(1+z)\,$K 
is the temperature of the CMB, $A_{10}=2.87\times 10^{-15}\,{\rm s^{-1}}$ is the spontaneous coefficient for the transition, $n_{\rm HI}$ is the proper 
neutral hydrogen density, and $H(z)$ is the Hubble parameter, $H(z)\simeq H_0\sqrt{\Omega_m}(1+z)^{3/2}$ at the redshifts of interest here.
Plugging in numbers from the \citet{planck16_PAR} 
base $\Lambda$CDM cosmology ($\Omega_mh^2=0.1417$, $\Omega_bh^2=0.0223$, $h=0.6774$, $Y_p=0.245$) gives

\begin{equation}
T_{\rm 21}(\nu_0) \simeq 26.9\,{\rm mK}~\left({1+z\over 10}\right)^{1/2}x_{\rm HI}\left(1-{T_{\rm CMB}\over T_S}\right), 
\end{equation}
where $x_\nHI$ is the globally-averaged neutral fraction. The signal will appear in absorption if $T_S<T_{\rm CMB}$ and emission 
otherwise. At the mean gas densities and low temperatures
corresponding to the absorption feature reported by the EDGES collaboration, spin-exchange collisions
are ineffective, and only the resonant scattering of ambient \Lya\ radiation  
can mix the hyperfine levels of the ground state and unlock the spin temperature from the CMB.
Assuming steady-state, the fractional deviation of the spin temperature from the temperature of the CMB is given by \citep{field58}

\begin{equation}
1-{T_{\rm CMB}\over T_S}={x_\alpha\over 1+x_\alpha}\left(1-{T_{\rm CMB}\over T_K}\right),
\label{eq:TS}
\end{equation}
where $x_\alpha$ is the \Lya\ coupling coefficients. Neutral hydrogen is therefore visible against the CMB only if the gas 
kinetic temperature differs from the CMB temperature and $x_\alpha>1$.\footnote{In principle, values of $x_\alpha<1$ may be sufficient 
to achieve a detectable absorption signal if the ratio between the radiation temperature and the gas temperature in Eq. (\ref{eq:TS}) is very 
large. In order to produce the best-fitting brightness temperature $T_{\rm 21}=-500\,$ mK observed at the center of the EDGES absorption trough 
with a coupling coefficient $x_\alpha<1$, this ratio would have to be larger than 27, compared to the value $T_{\rm CMB}/T_K=7$ expected in a standard
gas and radiation temperatures history. We shall not consider this extreme possibility further in this paper.}

\section{\Lya\ Coupling}

The Wouthuysen-Field mechanism \citep{wouthuysen52,field58} mixes the hyperfine levels of neutral hydrogen via the 
intermediate step of transitions to the $2p$ state and is key to the detectability of a 21-cm signal from the epoch of first light. 
In the cosmological context, we are principally interested in ``continuum" photons emitted by the first UV sources between the \Lya\ and \Lyb\ 
frequencies and redshifted into the \Lya\ resonance at $\nu_\alpha=2.47\times 10^{15}\,$Hz.
The coupling coefficient can be written as 

\begin{equation}
x_\alpha = {4\pi e^2 f_\alpha T_*\over 27 A_{10}T_{\rm CMB}m_e}\,S_\alpha n_{\alpha},
\\[8pt]
\end{equation}
where the factor $4/27$ relates the $1\rightarrow 0$ de-excitation rate via \Lya\ mixing to the total \Lya\ scattering rate \citep{field58}, 
$T_*\equiv h_P\nu_{10}/k_B=68.2$ mK, $f_\alpha=0.4162$ is the oscillator strength of the \Lya\ transition, and $n_{\alpha}$ is the specific 
photon number density per unit proper volume at the \Lya\ frequency in the absence of scattering (in units of $\nunits$).
The correction factor $S_\alpha\equiv \int d\nu n_\nu \phi_\nu/n_{\alpha}$, where $\phi_\nu$ is the line profile, accounts for spectral 
distorsions near the resonance \citep{chen04}. The condition $x_\alpha=1$ is satisfied when

\begin{equation}
n_{\alpha} = n_{\alpha}^c\equiv
4.2\times 10^{-20}{\rm cm^{-3}\,Hz^{-1}}\,\left({1+z\over 18}\right)\,S_\alpha^{-1}.
\label{eq:nuac}
\end{equation}
This ``thermalization" photon density corresponds to 
\begin{equation}
(n_{\alpha}^c\,\nu_\alpha/n_\nH) \simeq 0.1 [18/(1+z)]^{2}\,S_\alpha^{-1} 
\end{equation}
\Lya\ photons per hydrogen atom per unit logarithmic frequency interval. Once this value is excedeed, the Wouthuysen-Field effect turns on and 
drives $T_S\rightarrow T_K$. The strength of the Wouthuysen-Field coupling is weakened at low temperatures by atomic recoil, which leads to a distinct dip in the 
frequency distribution of photons near line center and pushes $S_\alpha$ below unity \citep[e.g.,][]{chen04,hirata06}.
To a good approximation the suppression of the scattering rate can be written as \citep{chuzhoy06}

\begin{equation}
S_\alpha=\exp(-0.013\tau_{\rm GP}^{1/3}/T_K^{2/3}),
\end{equation}
where $\tau_{\rm GP}\equiv \pi e^2 f_\alpha n_{\rm HI}/(m_eH\nu_\alpha)$ is the Gunn-Peterson optical depth and $T_K$ is in Kelvin.
At redshift 17 $\tau_{\rm GP}=1.59\times 10^6$ and the gas temperature is expected 
to be about 6.9\,K in the absence of astrophysical heating or non-standard cooling, so $S_\alpha\simeq 0.66$.\footnote{The above discussion has 
neglected a second correction to the Wouthuysen-Field coupling, 
whereby spin-exchange scatterings cause the photon spectrum to relax not to $T_K$ but to a temperature between $T_K$ and $T_S$ \citep{chuzhoy06,hirata06}.  
This is a small effect, leading to a $\lta 10\%$ reduction in the coefficient $x_\alpha$ for $T_K\gta 4\,$K.}~     


In the neutral IGM prior to reionization, photons originally emitted between the \Lya\ and \Lyb\ frequencies will redshift directly 
into the \Lya\ resonance, while those emitted at frequencies between \Lyg\ and the Lyman edge will redshift until they reach a Lyman-series 
resonance and excite a hydrogen atom into the $np$ configuration ($n\ge 4$). Because the gas is optically thick to \Lyn\ transitions, 
the excited atom will decay back to $1s$ in a radiative cascade that ultimately terminates either in a \Lya\ photon or in two $2s \rightarrow 1s$ 
photons. 
Quantum selection rules and Einstein $A$ coefficients determine the probability $P_{np}$ for an \HI\ atom 
in the $np$ configuration to generate a \Lya\ photon, $P_{np}=(1.000,0.000,0.261,0.308,0.326,...)$ for $n=(2,3,4,5,6,...)$ \citep{hirata06,pritchard06}. 

Let us now denote with ${\dot n}_{\nu'}(z')$ the average number of photons emitted per unit comoving volume per unit proper time per unit frequency at 
redshift $z'$ and frequency $\nu'$ by the first generation of UV sources. The resulting photon proper number density at redshift $z$ and 
frequency $\nu_\alpha$ can then be approximated in the limit of large optical depths as a weighted sum over the \Lyn\ levels 
\citep{barkana05,hirata06,pritchard06,meiksin10}

\begin{equation}
n_{\alpha}(z)= (1+z)^2 \sum\limits_{n=2}^\infty P_{np} \int_z^{z_{\rm max}(n)}\, {dz'} {{\dot n}_{\nu'}(z')\over H(z')},
\\[8pt]
\label{eq:ndot}
\end{equation}
where in each term $\nu'=\nu_n(1+z')/(1+z)$ is the emitted photon frequency at $z'$ corresponding to $1s\rightarrow np$ absorption at $z$,
$\nu_n=(4/3)\nu_\alpha(1-1/n^2)$ is the frequency of the \Lyn\ transition, $1+z_{\rm max}(n)=(1+z)\nu_{n+1}/\nu_n$ is the maximum redshift 
from which a photon entering the \Lyn\ resonance at $z$ can be observed.
Raman scattering in the Lyman series imposes a series of closely-spaced horizons, with the integral in each term of 
the sum in Equation ({\ref{eq:ndot}) being carried over redshift intervals, $\Delta z_n/(1+z)\equiv [z_{\rm max}(n)-z]/(1+z)$
%
%
which become increasingly smaller with higher levels.
Higher order cascades contribute then only a small fraction of the total \Lya\ flux. For photons originally emitted between the \Lya\ and \Lyb\ 
frequencies, one has $\Delta z_2/(1+z)=5/27$. 

\begin{figure}
\centering
\includegraphics[width=0.49\textwidth]{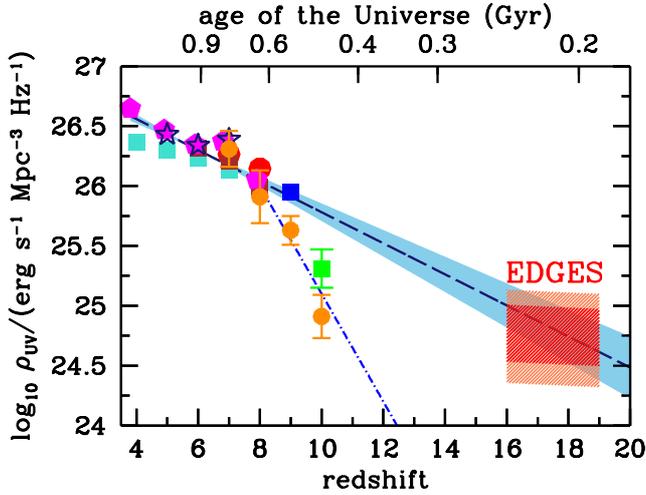}
\vspace{-0.6cm}
\caption{Constraints on the epoch of first light with the Wouthuysen-Field effect. 
The UV luminosity density inferred from the EDGES signal in the ``minimal coupling" regime ($1\le x_\alpha\le 3$) is plotted as the 
shaded red box at $16<z<19$, for an assumed $g$ factor of $0.06$.
The light shading reflects the estimated uncertainty in the $g$ parameter, $g=0.06\pm 0.02$ (see text for details).
The data points show the UV luminosity densities from galaxies using data from 
\citet{bouwens15LF} (magenta pentagons), 
\citet{bowler15} (blue stars),
\citet{finkelstein15} (turquoise squares),
\citet{ishigaki18} (orange circles),
\citet{livermore17} (brown squares),
\citet{mcleod15} (blue square), 
\citet{mclure13} (red circles),
and \citet{oesch18} (green square).
When computing luminosity densities, all luminosity functions have been integrated down to $M_{\rm lim} = -13$. 
The dashed line and pale blue shading depict our best-fitting slope for the growth of the galaxy UV luminosity density, 
$d\log_{10}\rho_{\rm UV}/dz=-0.130\pm 0.018$ (68\% confidence interval) in the redshift interval $4\le z\le 9$. 
%
%
The dot-dashed curve with $d\log_{10}\rho_{\rm UV}/dz=-0.45$ marks the accelerated evolution model of \citet{oesch18}, a
more rapid decline in the galaxy luminosity density at $z\gta 10$ that is inconsistent with the EDGES signal. 
}
\label{fig1}
\end{figure}

\section{Early Star-Formation}

Young, dust-free stellar populations with extremely low metallicities are characterized by very blue rest-frame UV continua, with spectral 
slopes around 1500\,\AA, $\beta\equiv d\ln f_\lambda/d\ln \lambda$, approaching the value $\beta\approx -3$ \citep{raiter10,schaerer03}. In 
this limit the source term ${\dot n}_{\nu}$  
does not depend on frequency, ${\dot n}_{\nu}\propto \nu^{-3-\beta}=\nu^0$. To account for redshift evolution effects over $\Delta z_n$, 
we write ${\dot n}_{\nu'}(z')=\dot n_{\rm UV}(z)E(z';z)$, where we have denoted the ratio of the comoving photon emissivity at redshift 
$z$ to that at redshift $z'>z$ as $E(z';z)$ [$E(z';z)=1$ in the case of no evolution]. In order to extract the main redshift dependence
of the \Lya\ photon flux in Equation ({\ref{eq:ndot}), it is convenient to define the following weighted integral 

\begin{equation}
{\bar E}_n(z)\equiv {\int_z^{z_{\rm max}(n)}\, {dz'} E(z';z)/H(z')\over \int_z^{z_{\rm max}(n)}\, {dz'}/H(z')}.
\end{equation}
The specific number of photons per unit proper volume at redshift $z$ and frequency $\nu_\alpha$ can then be written as 

\begin{equation}
n_{\alpha}(z) = {2(1+z)^{3/2}\over H_0\sqrt{\Omega_m}}  \dot n_{UV}(z)g
\label{eq:nua}
\end{equation}
where

\begin{equation}
g\equiv \sum\limits_{n=2}^\infty P_{np} {\bar E}_n(z)\left(1-\sqrt{\nu_n/\nu_{n+1}}\right).
\label{eq:g}
\end{equation}
We have folded a variety of parametrization for the redshift evolution law $E(z';z)$ into the $g$ term assuming a
a smoothly-varying monotonic function, and find typical values in the range $g\simeq 0.04-0.09$, with a contribution to the total \Lya\ photon 
flux from $4\le n\le 20$ \Lyn\ cascades that is typically $\lta 15$\%. In the case of a non-evolving emissivity and only counting photons 
originally emitted between the \Lya\ and \Lyb\ frequencies, one has $g=0.0814$. 

Substituting now for $n_{\alpha}$ in Equation (\ref{eq:nua}) the thermalization photon density $n_{\alpha}^c$ (Eq. {\ref{eq:nuac}) 
needed for efficient coupling gives the following constraint on the photon emissivity $\dot n_{UV}$,

\begin{equation}
\dot n_{UV}(z)> {H_0\sqrt{\Omega_m}\over (1+z)^{3/2}g}\,{27 A_{10}T_{\rm CMB}m_e\over 8\pi e^2 f_\alpha T_*\,S_\alpha}.
\end{equation}
Plugging in numbers and converting the photon emissivity $\dot n_{\rm UV}$ into a 1500\,\AA\ luminosity density for comparison with lower redshift data, 
$\rho_{\rm UV}=\dot n_{\rm UV}\,(h_P\nu_{1500})$ (where $\nu_{1500}=10^{15.3}\,$Hz), we finally derive

\begin{equation}
\rho_{\rm UV}(z)> 10^{24.52} g_{0.06}^{-1}\left({18\over 1+z}\right)^{1/2}\,\emunits, 
\end{equation}
where $g_{0.06}\equiv g/0.06$.
The numerical value on the rhs corresponds to the coupling condition $x_\alpha=1$. The best-case scenario for producing a strong 21-cm absorption signal 
is to assume $T_K\ll T_{\rm CMB}$ and $x_\alpha\gta 1$, and in the limit $T_K\ll T_{\rm CMB}x_\alpha$ Equation (\ref{eq:TS}) 
gives $T_S/T_K=(1+x_\alpha)/x_\alpha$. Let us then define the regime $1< x_\alpha < 3$ as one of ``minimal coupling", 
where the spin temperature approaches the gas temperature from above to within $1.33 < T_S/T_K\le 2$. 
This minimal coupling regime corresponds to a UV luminosity density in the range $\rho_{\rm UV}=10^{24.52}-10^{25.0}\,g^{-1}_{0.06}\,
[18/(1+z)]^{1/2}\,\emunits$. 

In Figure \ref{fig1} we have plotted this range as a shaded red box in the interval $16<z<19$, the approximate 
duration of the EDGES 21-cm absorption trough when efficient coupling must be maintained.
The figure also shows the UV luminosity density at $5\lta z\lta 10$ obtained by integrating the observed 
galaxy luminosity function down to a threshold $M_{\rm lim}=-13$. A turn-over in the $z\sim 7$ luminosity function at these faint 
magnitudes is inferred by combining the abundance matching technique with detailed studies of the
color-magnitude diagram of low-luminosity dwarfs in the Local Group \citep{boylan15}. 
Interestingly, the \Lya\ background flux inferred from the EDGES signal is entirely consistent with 
an extrapolation of UV measurements at lower redshifts, and does not require a substantially enhanced star formation or new exotic sources.
The function $\log_{10} (\rho_{\rm UV}/\emunits)=(26.30\pm 0.12) + (-0.130\pm 0.018)(z-6)$, depicted by the blue shading, is the 
our best fit for the growth of the galaxy UV luminosity density down to $M_{\rm lim}=-13$ mag and over the redshift interval 
$4\le z \le 9$. If our analysis is correct, it would appear that galaxy luminous mass built up at a remarkably steady rate over 
the first Gyr of cosmic history. Intringuingly, the same scaling also predicts inefficient Wouthuysen-Field coupling and therefore 
no signal at $z>20$, in agreement with the EDGES observations. As noted by \citep{kaurov18}, however, the sharpness of the brightness 
temperature drop between $z=21$ and $19$ suggests that the spin temperature of neutral hydrogen was coupled to the kinetic temperature 
of the gas very rapidly, within a small fraction of a Hubble time.
This requires the \Lya\ background to decline at $z>20$ more abruptly than expected from the $d\log_{10}\rho_{\rm UV}/dz=-0.130$ evolution.

Figure \ref{fig1} also displays (dot-dashed curve with $d\log_{10}\rho_{\rm UV}/dz=-0.45$) the accelerated evolution suggested by the dearth 
of galaxies brighter than $-17$ mag at $z\sim 10$ in {\it Hubble Space Telescope (HST)} deep fields \citep{oesch18,ishigaki18}. 
Based on a comprehensive search in all prime {\it HST} datasets, \citet{oesch18} have recently shown that the UV luminosity function decreases by one 
order of magnitude from $z\sim 8$ to $z\sim 10$ over a four magnitude range, a drop that may signal a shift of star formation toward less massive, fainter galaxies.
The lower normalization -- if not accompanied by a steepening of the faint-end slope with redshift -- implies a rapid decrease of the 
total UV luminosity density at these epochs, a decline that extrapolated to $z\gta 15$ is clearly inconsistent with the EDGES signal. The debate
about the UV luminosity function at $z\sim 10$ is far from settled, however, and the {\it HST} and EDGES observations may hint at a 
possible differential evolution of bright vs. faint galaxies \citep{mason15,mirocha18}.

\section{Ionizing Photon Production}

Strickly speaking, of course, the 21-cm coupling constraints plotted in Figure \ref{fig1} should be considered only as a lower limit to the UV 
luminosity density at $16<z<19$. It is instructive, at this stage, to estimate the emission above the Lyman limit expected by the same galaxies that induce strong 
\Lya\ coupling. Population synthesis models predict an ionizing photon production efficiency per unit stellar mass that increases with decreasing metallicity
and for initial mass functions (IMFs) favouring the formation of very massive stars \citep{schaerer03}. 
Here, we are interested in the LyC output per unit UV luminosity density, $\xi_{\rm ion}\equiv \int_{\nu_L} d\nu {\dot n}_\nu/\rho_{\rm UV}$, 
where $\nu_L$ is the hydrogen Lyman edge. For extremely metal-poor as well as young (zero age main sequence) stellar populations, the ionizing photon yield 
is only weakly dependent on the IMF and converges to the narrow range of values $\log_{10} (\xi_{\rm ion}/\xiunits) = 26-26.2$ \citep{raiter10}. 
Constant star-formation rate models reach an equilibrium value of the yield $\xi_{\rm ion}$ that is 0.1-0.15 dex smaller in the case of massive IMFs, 
and even smaller for ``normal" IMFs. These yields are consistent with those estimated from H$\alpha$ emission line fluxes in the bluest and faintest galaxies 
at $5.1<z<5.4$, $\log_{10} (\xi_{\rm ion}/\xiunits)=25.9^{+0.4}_{-0.2}$ \citep{bouwens16xion}.

Extremely metal-poor stellar systems with a UV luminosity density of $\rho_{\rm UV}=10^{24.82}\,g^{-1}_{0.06}[18/(1+z)]^{1/2}\,{\rm erg\,s^{-1}\, Mpc^{-3}\,Hz^{-1}}$, 
corresponding to a coupling coefficient $x_\alpha=2$, will then produce somewhere in the range of 

\begin{equation}
{\xi_{\rm ion}\rho_{\rm UV}\over n_{\rm H}}\simeq 0.3-0.6\, g^{-1}_{0.06}\,\left({18\over 1+z}\right)^{1/2}\,\left({x_\alpha\over 2}\right)
\label{eq:xi}
\end{equation}
ionizing photons per H-atom per 100 Myr, where the scaling with coupling strength has been made explicit for clarity. 
Of these, only a globally-averaged fraction $f_{\rm esc}$ will escape from individual galaxies, make it into the IGM, and 
initiate the process of reionization by creating expanding \HII\ bubbles in the neutral cosmic gas. 
Since photoionizations dominate over radiative recombinations at these redshifts, the ``reionization equation" for the time evolution of 
the volume-averaged hydrogen ionized fraction,

\begin{equation}
{dx_{\rm HII}\over dt}=f_{\rm esc}\,{\xi_{\rm ion}\rho_{\rm UV}\over n_{\rm H}}, 
\\[8pt]
\label{eq:xHII}
\end{equation}
can be easily integrated. The escape fraction is constrained by the latest {\it Planck} CMB anisotropy and polarization data analysis, which 
yields a small Thomson scattering optical depth of $\tau_{\rm es}\simeq 0.055-0.060$, with $\sigma(\tau_{\rm es})\sim 0.01$ \citep{planck16_REI}; a 
redshift-asymmetric parameterization of the ionized fraction gives $z_{\rm beg}=10.4^{+1.9}_{-1.6}$ for the redshift (``beginning" of reionization) 
where $x_{\rm HII}=10$\%, and leaves little room for any significant ionization at $z>15$.
More specifically, a recent non-parametric reconstruction of the history of reionization applied to {\it Planck} intermediate 2016 data 
finds a 68\% upper limit to the electron fraction at $z=15$ of 8\% (flat $\tau_{\rm es}$ prior, see \citealt{millea18}).

A fiducial model with $\xi_{\rm ion}\rho_{\rm UV}/n_{\rm H}=0.5 [18/(1+z)]^{1/2}$ ionizing photons per H-atom per 100 Myr would require then, according 
to Equation (\ref{eq:xHII}), mean escape fractions $f_{\rm esc}\lta 20\%$ to keep the early IGM largely neutral and delay the reionization process
consistently with {\it Planck} results. Any early stellar systems producing a larger \Lya\ background intensity would emit more LyC
and X-ray radiation, generating more ionizations and gas heating which would tend to make the depth and duration of the absorption signal smaller. 
For comparison, studies of the reionization history at later epochs have shown 
that, when assuming a limiting magnitude of $M_{\rm lim}=-13$ and $\log_{10} (\xi_{\rm ion}/\xiunits) = 25.2-25.5$, average escape fractions from galaxies
of $10-20$\% produce the requisite number of LyC radiation to complete reionization by $z=6$ 
\citep[e.g.,][]{bouwens15rei,finkelstein15,robertson15,madau17} with no contribution from other sources.

\section{Summary}

The 21-cm global signal is a window into the earliest star formation in the universe. In this study we have ignored the anomalous depth of the EDGES 
absorption feature, and concentrated instead on the 
implications for early star formation of the \Lya\ photon field required to couple the spin temperature of the hyperfine levels to the gas temperature.
Under the basic assumption that the 21-cm signal is activated by young, metal-poor stellar systems, we have shown that the EDGES signal is consistent with 
an extrapolation of the evolving galaxy UV luminosity density measured at $4\lta z\lta 9$ by deep {\it HST} observations. Models of accelerated evolution 
where the UV luminosity density declines rapidly at $z\gta 10$ are unable to provide the needed \Lya\ coupling strength at $z\lta 20$. If our analysis is 
correct and the EDGES results are confirmed, galaxy light appears to have built up at a surprisingly steady rate over the first Gyr of cosmic history; in other words, 
a substantially enhanced star formation rate density or exotic luminous sources do not seem to characterize the epoch of first light. The amount of 
ionizing radiation expected by the same systems that induce strong \Lya\ coupling is significant, of order 0.3--0.6 LyC photons per H-atom per 100 Myr 
at these epochs. To keep the early IGM largely neutral and delay the reionization process consistently with {\it Planck} CMB 
results, our estimates imply mean escape fractions into the IGM $f_{\rm esc}\lta 20$\%. 

Since we have focused on the evolution of the cosmic UV emissivity, the constraints discussed here -- while still uncertain -- are rather model 
independent, as we have tried to keep 
assumptions regarding the nature of the first luminous sources to a minimum. The present analysis should be regarded as complementary to those of, e.g., 
\citet{mirocha18} and \citet{kaurov18}, who have recently explored the consequences of the EDGES results on the star formation efficiencies of dark matter 
halos at early times. We finally note that, in the presence of some other 21-cm background 
of specific intensity $J_{21}$, the brightness temperature of such radio emission should be added to $T_{\rm CMB}$ in all the above equations, 
$T_{\rm CMB}\rightarrow T_R=T_{\rm CMB}+c^2J_{21}/(2k_B\nu_{10}^2)$. A radio background excess \citep{ewall-wice18,feng18}
would require an even larger \Lya\ radiation field in order to achieve efficient coupling.

\section*{Acknowledgments} \ni
The author would like to thank R. Bouwens and A. Meiksin for many useful discussions on the topics presented here. 
Support for this work was provided by NASA through a contract to the WFIRST-EXPO Science Investigation Team (15-WFIRST15-0004), administered by the GSFC.

\bibliographystyle{mnras}
\bibliography{paper}

\bsp	
\label{lastpage}
\end{document}